\documentclass[aps,prl,twocolumn,superscriptaddress]{revtex4-1}
\usepackage{graphicx}
\usepackage{dcolumn} 
\usepackage{graphics}
\usepackage{amsmath,amsfonts,amssymb}
\usepackage[colorlinks=true,allcolors=blue]{hyperref}
\usepackage{float}

\usepackage{xspace} 
\usepackage[svgnames]{xcolor}
\usepackage{cancel}
\usepackage{bbold}

\newcommand{\eq}[1]{\begin{equation} #1 \end{equation}}

\begin{document}

\title{Josephson Junctions Via Anodization of Epitaxial Al on an InAs Heterostructure}

\author{A.~Jouan}
\affiliation{ARC Centre of Excellence for Engineered Quantum Systems, School of Physics, The University of Sydney, Sydney, NSW 2006, Australia}

\author{J.~D.~S.~Witt}
\affiliation{ARC Centre of Excellence for Engineered Quantum Systems, School of Physics, The University of Sydney, Sydney, NSW 2006, Australia}

\author{G.~C.~Gardner}
\affiliation{Microsoft Quantum Purdue, Purdue University, West Lafayette, Indiana, USA}

\author{C.~Thomas} 
\affiliation{Microsoft Quantum Purdue, Purdue University, West Lafayette, Indiana, USA}
\affiliation{Department of Physics and Astronomy, Purdue University, West Lafayette, Indiana, USA}

\author{T.~Lindemann} 
\affiliation{Microsoft Quantum Purdue, Purdue University, West Lafayette, Indiana, USA}
\affiliation{Department of Physics and Astronomy, Purdue University, West Lafayette, Indiana, USA}

\author{S.~Gronin} 
\affiliation{Microsoft Quantum Purdue, Purdue University, West Lafayette, Indiana, USA}

\author{M.~J.~Manfra}
\affiliation{Microsoft Quantum Purdue, Purdue University, West Lafayette, Indiana, USA}
\affiliation{Department of Physics and Astronomy, Purdue University, West Lafayette, Indiana, USA}

\author{D.~J.~Reilly}
\affiliation{ARC Centre of Excellence for Engineered Quantum Systems, School of Physics, The University of Sydney, Sydney, NSW 2006, Australia}
\affiliation{Microsoft Quantum Sydney, The University of Sydney, Sydney, NSW 2006, Australia}

\date{\today}

\begin{abstract}

We combine electron beam lithography and masked anodization of epitaxial aluminium to define tunnel junctions via selective oxidation, alleviating the need for wet-etch processing or direct deposition of dielectric materials. Applying this technique to define Josephson junctions in proximity-induced superconducting Al-InAs heterostructures, we observe multiple Andreev reflections in transport experiments, indicative of a high quality junction. We further compare the mobility and density of Hall-bars defined via wet etching and anodization. These results may find utility in uncovering new fabrication approaches to junction-based qubit platforms.
\end{abstract}

\maketitle

Zero energy modes in a 1-dimensional (1D) topological superconductor are candidate building blocks for high-fidelity quantum computation \cite{Kitaev2001, Nayak2008}. One proposed approach to realising a topological superconductor is via a 2-dimensional electron gas (2DEG) with large spin-orbit interaction, proximity coupled to a conventional superconductor and confined to 1D using electrostatic gates and tunnel barriers \cite{Lutchyn2010,Oreg2010, Mourik2012}. Such a configuration is experimentally challenging to realize however, since many of the needed parameters also compete with each other. Inducing superconductivity \cite{Kjaergaard2016}, for instance, requires that the 2DEG be close to the superconductor, where disorder from surface states can also greatly degrade the mobility \cite{Hatke2017}. 

\begin{figure}
	\includegraphics[width=1.0\columnwidth]{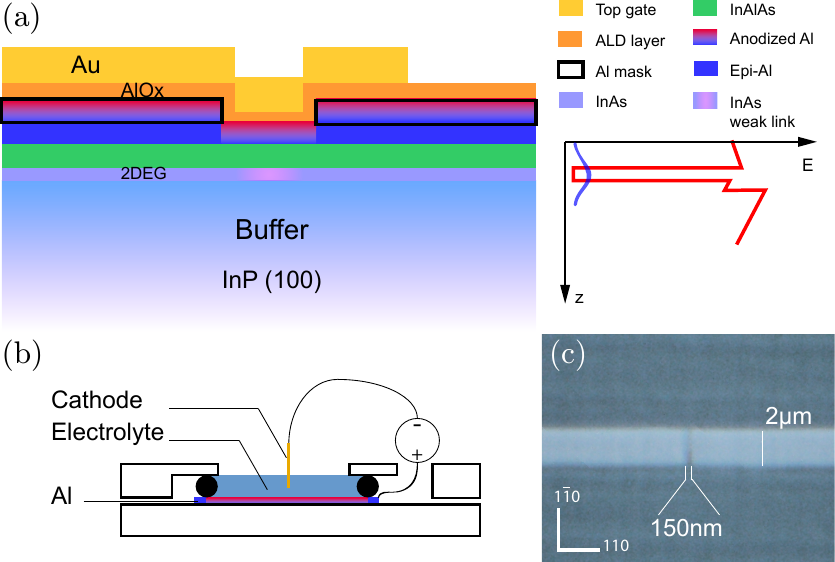}
	\caption{(a)  InAlAs/InAs/InGaAs quantum well covered with an \emph{in situ} grown 7~nm epi-Al layer. An Al-mask (circled in black) was patterned using electron beam lithography. Finally, the sample was anodized and covered with ALD-grown Al$_2$O$_3$ and Au top gate. (inset) A schematic representation of the quantum well (in red) and the wave function of the first filled level. (b) Schematic of the jig used to anodize the Al. A mixture of ammonium pentaborate tetra-hydrate with ethylene glycol, following ref. \cite{Kroger1981}, is poured into the jig and a current of $50$~$\mu$A flows between the needle and surface of the sample. (c) Optical microscope image of the 2~$\mu$m wide planar Josephson junction after the ALD-grown Al$_2$O$_3$ layer was deposited. The junction was aligned along the 110 orientation of the InAs heterostructure beneath. It is 2~$\mu$m wide, surrounded by normal 2DEG regions, and has a gap of 150~nm.}
	\label{anod}
\end{figure} 

Josephson junctions or tunnel barriers separating topological segments from normal regions are particularly sensitive to disorder. Typically, crystaline superconductors, grown using molecular beam epitaxy (MBE) are etched away using invasive wet chemical processing or damaging dry etchant techniques \cite{Richter2002,Heida1998,Mur1996}. This approach is often used to create, for example, the weak link of a Josephson junction via wet etching of Al (usually with transene) \cite{Fornieri2019}.  Although there is opportunity to further  optimize the etching process \cite{Pauka2020}, circumventing the need for it entirely would likely lead to junctions with less disorder. An associated challenge is the patterning of features on the length-scale of a few tens of nanometers in device structures, for example, to be able to reach the Josephson short-junction limit \cite{Peng2016}. At these length-scales wet-etching can be difficult to control.

Here, we report the use of anodization to define a Josephson junction via the selective oxidation of epitaxial aluminium on an InAs heterostructure. In addition to being largely non-invasive, the technique offers a means of defining structures with a resolution limited, in-principle,  only by the nanometer precision of the electron beam lithography. In a straight up comparison we also show that Hall bars defined via anodization have a moderately higher mobility than devices that are shaped by wet-etch processing.


Anodization is a standard industrial method used extensively to protect or modify the properties of surfaces \cite{Diggle1969}. It has a wide range of applications, most abundantly for protecting metallic objects from uncontrolled oxidation. In the field of device physics, it has been used to define Nb Josephson junctions \cite{Kroger1981}.


The anodization circuit is realized such that the target metal constitutes the anode. By driving a current through the target metal submerged in a suitable electrolyte, an oxydo-reduction reaction can be induced. In the case of Al, alumina ($Al_2O_3$) is formed:
\eq{2Al(s) + 3H_2O(l) = Al_2O_3(s) + 6H^+ + 6e^-} 
This process allows for the controlled oxidation of Al to proceed to a specific depth beyond the 3~nm native oxide \cite{Mott1947}, at a rate fixed by the applied current \cite{Hunter1954}.

Turning to quantum devices, we make use of this reaction to define tunnel barriers in materials comprising superconducting heterostructures. Unlike conventional fabrication approaches that remove the Al superconductor and back-fill with dielectric, here we rather use a second lift-off patterned layer of evaporated Al (Al-mask) on top of the epi-Al and employ the anodization process. In this way, we are able to retain the high-resolution pattern of the evaporated Al-mask, and, also the high-quality interface between the heterostructure and the newly formed dielectric near the shallow quantum well.

\begin{figure}
	\includegraphics[width=1.0\columnwidth]{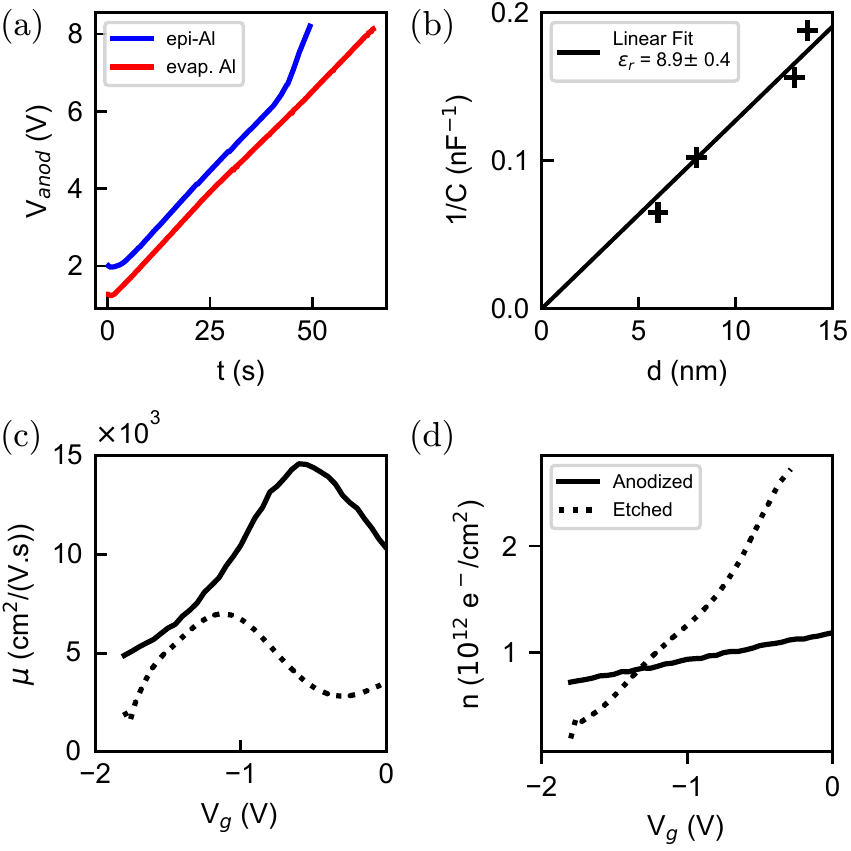}
	\caption{(a) Anodization voltage, $V_{anod}$, measured for the two different type of Al. In blue, the epi-Al on an InAs quantum well, where a clear kink above 6 V corresponds to the point at which all the Al has been anodized. In red, the anodization of a 10~nm thick evaporated Al-mask. The different initial voltage correspond to different thicknesses of the native oxide layer. A voltage of $V_{anod}(t=0) = 2.1$ V corresponds to a thickness of 3~nm of alumina. (b) Fit of the capacitance of squares of 1~mm by 1~mm as a function of the thickness of anodized alumina. The capacitance was obtained by measuring the out of phase response of a 4~mV applied voltage at $f = 30$ Hz using a lock-in. (c) Mobility, $\mu$ as a function of top-gate voltage measured in a Hall bar made either with the same Al-mask anodization technique or with an etched method from the same wafer (d) Electron density, $n$, as a function of top-gate voltage extracted from the Hall measurement for anodized and etched samples.}
	\label{V_t_compare}
\end{figure} 

Starting from an InAlAs/InAs/InGaAs heterostructure (quantum well 10~nm below the surface and 7~nm of \emph{in situ} grown epi-Al),  \cite{Hatke2017}, a poly(methyl methacrylate) resist is spin-coated and patterned using electron beam lithography, as shown in Fig.~\ref{anod}~(a). The Al-mask layer -- the same thickness as the epi-Al layer -- is then evaporated and the pattern formed using the lift-off technique. Immediately prior to the evaporation of the Al-mask layer, in the same vaccum cycle, an Ar ion mill step is performed to remove the native oxide from the epi-Al, ensuring good electrical contact between the two Al layers. The sample is then placed in a setup as shown in Fig.~\ref{anod}~(b), submerged in an electrolyte -- a mixture of ammonium pentaborate tetra-hydrate and ethylene glycol \cite{Kroger1981} -- and connected to the positive terminal of the current source. Thereby the surface Al constitutes the anode. A needle is used as the cathode,  and is placed into the electrolyte solution a small distance from the sample. The structure is anodized using a current of 50 $\mu$A and up to 6 V [Fig.~\ref{anod}~(c)]. While a fixed current is passing through the anodization circuit, a voltage $V_{anod}$, exists across the developing oxide (thickness, $d$). This voltage is fixed by the critical electric field, $E_c = V_{anod}/d$, needed to force the ions through the oxide layer \cite{Diggle1969}. This voltage increases linearly with time during anodization and allows us to monitor the thickness of the oxide layer and the rate of anodization, as shown in  Fig.~\ref{V_t_compare}~(a). By monitoring $V_{anod}$ as a function of time, the effective depth of the anodized Al can be determined. 
 
The anodization is stopped at a known voltage, $V_{anod}$, which corresponds to the thickness of the Al-mask. In this way, the Al-mask has effectively been anodized instead of the epi-Al in the specifically patterned regions. The regions that were not covered by the Al-mask are now fully insulating, while the parts of the epi-Al which where covered with the Al-mask are now covered with a layer of alumina. An anodization rate of 100 mV/s for both the Al-mask and epi-Al has been measured. It is worth emphasizing that for this method to work the rates for both must be equal. This ensures that the Al-mask and the uncovered epi-Al layer will be fully anodized at the same point in time.

Following the anodization, a mesa structure is etched in order to avoid top-gate leakage. The alumina around the mesa is removed using a developing solution, AZ726, followed by a dilute phosphoric acid etch. Finally, a 5 nm thick alumina layer is grown by ALD and the Ti/Au top-gates are deposited, as shown in Fig.~\ref{anod}~(a).


The anodization ratio of oxide thickness, $d$, to measured voltage, $V_{anod}$, was determined by performing ellipsometry on test samples anodized to different depths. A ratio of 1.37~$\text{nm}.\text{V}^{-1}$ was found, which is consistent with values in the literature \cite{Diggle1969, Hunter1954}. A further corroboration of the extrapolated rate is the clear kink in $V_{anod}$ vs.~$t$ of the epi-Al anodization, which can be seen in Fig.~\ref{V_t_compare}~(a). This voltage translates to an epi-Al thickness of 7~nm, which matches that measured by X-ray diffraction. The change of rate of anodization corresponds to a higher critical electric field, $E_c$, needed to force ions through the InAlAs layer. Consistent with this picture we estimate the dielectric constant of the oxide by measuring the capacitance, $C = \epsilon_0 \epsilon_r S/d$, between two patterned pads of surface, $S$, with different anodized alumina of known thickness, $d$, see Fig.~\ref{V_t_compare}~(b). The relative permeability was found to be, $\epsilon_r = 8.9 \pm 0.4$.

\begin{figure}
	\includegraphics[width=1.0\columnwidth]{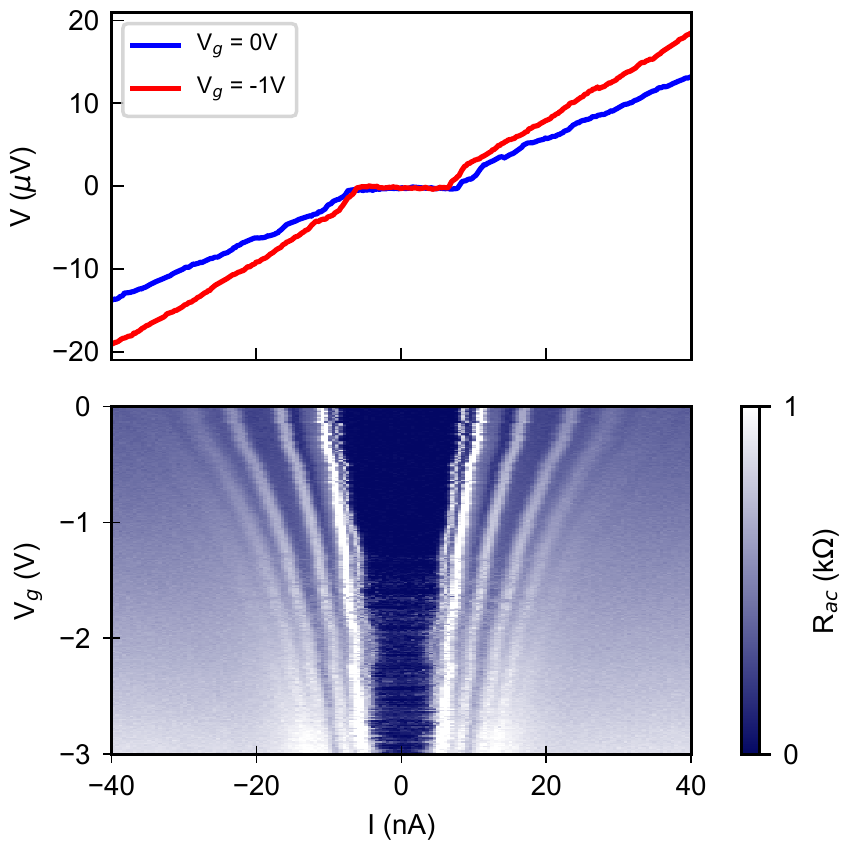}
	\caption{Upper pannel: I-V characteristic of the Josephson junction formed by anodization. The junction is current biased with a 1~MOhm resistor. Two I(V) at different gate voltage are shown. Kinks in the I-V characteristic can be seen around 10~$\mu$V and below. Lower panel: differential resistance versus bias current as a function of top gate voltage. The excitation frequency, $f = 77.78$~Hz, and an amplitude of 0.2~mV across a 1~MOhm resistor, which corresponds to a current of $I_{ac} = 200$~pA.}
	\label{IV_GV}
\end{figure} 


Next, we compare the mobility and electron density of Hall bars defined using standard wet etching to those defined by anodization, as shown in Fig.~\ref{V_t_compare}~(c) and Fig.~\ref{V_t_compare}~(d). Here, data is taken at an electron temperature of $T_e = 55$~mK. For the Hall-bar defined using anodization, the mobility at ($V_g=0V$) was found to be $\mu \sim  12,000$~cm$^2$/(V.s), with electron density ($V_g=0V$) $n = 1.2 \times 10^{12}$ e$^-$/\text{cm}$^2$, comparable to values given in the literature \cite{Tschirky}. Application of a negative gate voltage leads to a peak mobility of $\mu \sim  15,000$~cm$^2$/(V.s) The variation in mobility  as a function of applied top-gate voltage is consistent with screening of charge impurities by electrons in the 2DEG \cite{Umansky1997}. 

For comparison, a sample from the same wafer was patterned into a Hall bar using an optimised standard Transene type-D etch \cite{Pauka2020} (otherwise processed identically) and was found to have a peak mobility of $\mu \sim  7,000$~cm$^2$/(V.s). We underscore the improvement in mobility of the device fabricated using anodization, indicating the sensitivity of these device structures to disorder induced by chemical processing. Also of note, for the anodized Hall bar, the electron density at $V_g= 0V$ is around half that of the chemically etched device. A similar anodization method has also recently yielded an enhancement of the mobility \cite{Drachmann2021}.



Finally, to demonstrate the utility of this technique for nanoscale device fabrication, a Josephson junction with a barrier region of length $\sim$~150~nm was fabricated using anodization, as shown in Fig.~\ref{anod}~(c). The device was characterized using a standard dc transport making use of a current-bias with four-point measurement in a dilution refrigerator with an electron temperature of $T_e = 55$~mK -- (determined using an NIS junction \cite{Feshchenko2015}). The epi-Al layer has a critical temperature above $T \sim 1.5$~K and an in-plane critical magnetic field, $B_c \sim 2.5$~T \cite{Suominen2017a}. The mobility obtained from the Hall measurements corresponds to an elastic scattering length, $l_e \sim 200$~nm, which suggests the junction is in the quasi-ballistic limit. 

The current--voltage characteristics of the device as a function of applied top-gate voltage can be seen in Fig.~\ref{IV_GV}. Despite the relatively low critical current compared to similar devices in the literature \cite{Mayer2019}, multiple Andreev reflections (MAR) are clearly resolved which indicates that the electrons remain phase coherent over multiple times the length of the weak link.

With increasing negative applied top-gate voltage, the 2DEG is depleted and the critical current decreases, as shown in  the lower panel of Fig.~\ref{IV_GV}. Below $V_g = -1.5$ V, the junction loses phase coherence and a finite voltage appears across the weak link. The characteristic voltage or $I_{c}R_{n}$ product is constant between 0~V and -1.5~V (see SI) at around 3~$\mu$eV, which is a few percent of the Al gap, $\Delta_{Al} = 1.76 k_{B}T_{c} \simeq 230$~$\mu$V. As an indication, the excess current at -1~V is measurable and is $I_{ex} \sim 0.95$~nA, which is approximately 15\% of the critical current. Although lower than the value expected for a fully transparent interface \cite{Kjergaard2017,Blonder}, further investigations into the exact nature of the induced superconducting gap, $\Delta^*$, in this particular type of heterostructure need to be performed in order to fully understand this discrepancy.

\begin{figure}
	\includegraphics[width=1.0\columnwidth]{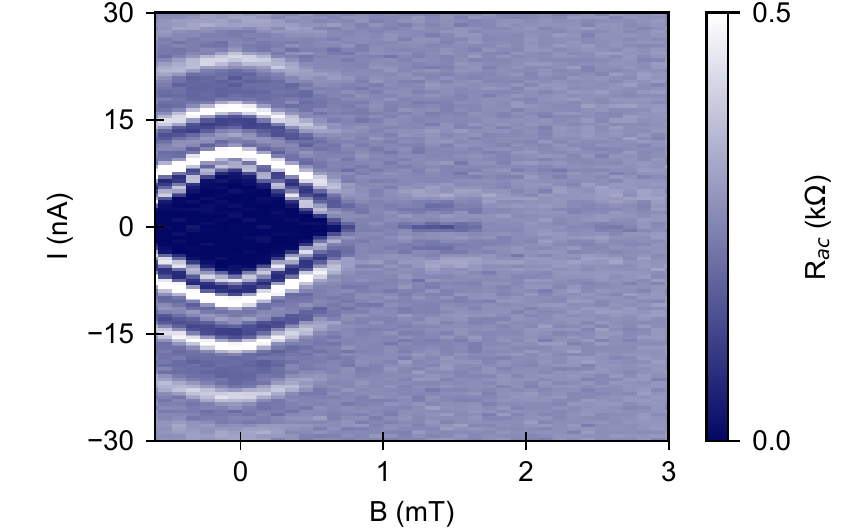}
	\caption{Differential resistance as a function of magnetic field and bias current at $V_g = 0$~V. Clear oscillation of the critical current can be observed as a function of the applied magnetic field perpendicular to the interface. The critical current goes to zero when the flux through the weak link corresponds to one flux quantum. Same settings as Fig.~\ref{IV_GV}~(Lower).}
	\label{B_field_sweep}
\end{figure}

Finally, we apply a perpendicular magnetic field and investigate the differential resistance of the junction, as shown in Fig.~\ref{B_field_sweep}. Fraunhoffer patterns are evident -- with some asymmetry between positive and negative applied bias. The critical current goes to zero twice in the range shown (three lobes) at the field values where an integer multiple of the flux quantum threads through the weak link \cite{Tinkham2004}. The measured field at which the critical current is cancelled by currents of opposite direction due to the flux through the junction is $\sim 1$~mT, which is lower than the expected value given the dimensions of the junction.  
This discrepancy can be explained either by magnetic flux focusing due to the Meissner effect in the superconducting leads or alternatively by an increased London penetration depth in the thin-film epi-Al. Such variation in the length-scales in thin-film superconductors is not uncommon, having been observed elsewhere \cite{Suominen2017a} and, indeed, are an area of active theoretical research. In conclusion, we have demonstrated that Al-mask anodization represents a viable route to nanoscale device fabrication, in particular  gate-tunable Josephson junctions in InAs quantum wells. The technique is versatile and likely applicable to various materials.

We thank S. Pauka and M. Cassidy for useful conversations and technical help. This research was supported by the Microsoft Corporation and the Australian Research Council Centre of Excellence for Engineered Quantum Systems (EQUS, CE170100009). The authors acknowledge the facilities as well as the scientific and technical assistance of the Research \& Prototype Foundry Core Research Facility at the University of Sydney, part of the Australian National Fabrication Facility.

\bibliographystyle{apsrev4-1}
%
\end{document}